\documentclass[11pt]{article}

\let\ssection=\section
\renewcommand{\section}{\setcounter{equation}{0}\ssection}

\newcommand\half{{\scriptstyle{\frac{1}{2}}}}
\def\parag{\hfil\break} 
\def\kikezd{\parag\underbar}
\def\smallcirc{{\raise 0.5pt \hbox{$\scriptstyle\circ$}}}
\def\p{{\partial}}
\def\vb{{\vec b}}
\def\vc{{\vec c}}

\def\vj{{\vec\jmath}}
\def\vp{{\vec p}}

\def\vj{{\vec\jmath}}

\def\vx{{\vec x}}

\def\vnabla{{\vec\nabla}}

\begin{document}

\setlength{\baselineskip}{16pt}

\title{Exotic galilean symmetry in non commutative field theory}

\author{
P.~A.~Horv\'athy
\\
Laboratoire de Math\'ematiques et de Physique Th\'eorique\\
Universit\'e de Tours\\
Parc de Grandmont\\
F-37 200 TOURS (France)
\\ and\\
L. Martina\\
Dipartimento di Fisica dell'Universit\`a
\\
and\\
Sezione INFN di Lecce. Via Arnesano, CP. 193\\
I-73 100 LECCE (Italy).
}

\date{\today}

\maketitle

\begin{abstract}
The non-relativistic version of the non commutative Field Theory,
recently introduced by Lozano, Moreno and Schaposnik [1],
is shown to admit the ``exotic'' Galilean  symmetry found before
for point particles.
\end{abstract}
\vskip5mm
\centerline{\texttt{hep-th/0207118}. (with Note added)}
\vskip5mm

\section{Introduction}

It has been known for some time that the planar Galilei group
admits a
two--parameter central extension \cite{LeLe,BGO,BGGK,GRIG,LSZ}.
The only physical examples with such an ``exotic'' Galilean symmetry
known so far are the scalar model in \cite{GRIG, DH}
equivalent to non-commutative
quantum mechanics (NCQM) \cite{NCQM} and used in the context of the
Hall Effect \cite{DH},
and the accele\-ration-dependent system in \cite{LSZ}.

Here we first revisit NCQM, viewed as a classical field theory.
Then we turn to non-commutative field theory (NCFT) \cite{NCfield}.
In these theories, which have attracted a considerable amount of
recent attention, the ordinary product
is replaced by the Moyal ``star'' product associated
with the posited non-commutative structure of the plane.
In \cite{LMS} Lozano et al. present in particular
a non-relativistic version of NCFT.
Below we point out that the  latter
theory admits an ``exotic'' Galilean symmetry
analogous to that of a point particle \cite{DH}.
\goodbreak

\section{NC Quantum Mechanics}\label{NCQM}

Let us consider a free scalar particle in the
non-commutative plane, given by
the standard hamiltonian $h=\vp^2/2m$ and
the fundamental commutation relations \cite{BGO,BGGK,GRIG}
\begin{equation}
\begin{array}{lll}
	\big\{x_{1},x_{2}\big\}=\theta,
	\\
	\big\{x_{i},p_{j}\big\}=\delta_{ij},
	\\
	\big\{p_{1},p_{2}\big\}=0,
\end{array}
\label{freeNCcomrel}
\end{equation}
where $\theta$ is the non-commutative parameter.
  As shown in Ref. \cite{DH},
the ``exotic'' [meaning two-fold centrally extended] Galilei group
is a symmetry for the system
with associated conserved quantities
$\vp$, $h$, $m$, $k=-m^2\theta$, augmented with
the modified angular momentum and Galilean boosts,
\begin{equation}
\begin{array}{ll}
\jmath&=
\displaystyle
\vx\times\vp+\frac{1}{2}\theta\vp{\,}^2+s,
\\[8pt]
g_{i}&=mx_i-p_it+m\theta\,\epsilon_{ij\,}p_{j},
\end{array}
\label{momentmap}
\end{equation}
where $s$ represents the anyonic spin.
The commutation relations of this algebra w. r. t. the ``exotic''
Poisson bracket (\ref{freeNCcomrel})
coincide with those of the ordinary, singly-extended
Galilei group except for the boosts, whose bracket
yields the second, ``exotic parameter''
\begin{equation}
\{g_{1},g_{2}\}=-m^2\theta\equiv -k.
\label{boostcr}
\end{equation}

Owing to the non-commutativity of the coordinates, the system has no
position representation: $x, y$ do {\it not} form a complete
{\it commuting} system. Put in another way: $\vx=$const. is not
a polarization\footnote{Quantization can be performed using
canonical [Darboux] variables; but these latter will not have
the physical interpretation of position, cf. the Discussion.}.
The momentum representation is still valid, though,
and we represent therefore the wave functions by
square-integrable functions of the momentum, $\phi(\vp)$.
The representation of the ``exotic''
Galilei group is given as \cite{BGO,BGGK,GRIG,DH}
\begin{equation}
U_{a}\phi(\vp)=\exp\left(
i\left[\frac{\vp{\,}^2e}{2m}-\vp\cdot\vc+s\varphi
  +m\,u\right]
+
im\theta\left[\frac{1}{2}\vb\times\vp+m\,v\right]\right)
\phi\left(R^{-1}(\vp-m\vb\,)\right),
\label{exoticrep}
\end{equation}
where $a$ is an element of the ``exotic'' Galilei group
with $e$ representing a time translation, $\vc$ a space translation,
$\vb$ the boosts, $R$ a rotation with angle $\varphi$;
  $u$ and $v$ represent the
translations along the central directions.

The infinitesimal action of (\ref{exoticrep}) yields the quantum operators.
The momentum operator is standard, $\widehat{p_i}$ is multiplication by
$p_i$; the energy is $\hat{h}=\frac{\vp^{2}}{2m}$.
For the other ``exotic'' Galilei generators we get
\begin{equation}
     \begin{array}{c}
	\hat{\jmath}=i\epsilon_{ij}p_{i}\displaystyle\frac{\p}{\p p_{j}}+s
	\hfill
	\\[6pt]
	\hat{g}_j=
	m\left(i\displaystyle\frac{\ \p }{\p p_{j}}
	+\displaystyle\frac{1}{2}\theta\epsilon_{jk}\,p_{k}\right)\qquad
	\hfill
	\end{array}
\label{exGalop}
\end{equation}
while $m$ and $k=-m^2\theta$ act trivially.
The particle satisfies the usual free  Schr\"odinger equation in
the momentum space,
\begin{equation}
     i\p_{t}\phi(\vp)=\frac{\vp^{2}}{2m}\phi(\vp).
     \label{freeschreq}
\end{equation}

The point is that QM  can be also viewed as a
  classical field theory. Eq. (\ref{freeschreq})
derives indeed from the Lagrangian
\begin{equation}
     L=\int\left(\frac{i}{2}\big(\bar{\phi}\p_{t}\phi-\phi\p_{t}\bar{\phi}\big)
     -
     \frac{\vp^2}{2m}\vert\phi\vert^2\right) d^2\vp.
     \label{freepLag}
\end{equation}

The theory given by (\ref{freepLag})
is manifestly invariant w. r. t. the ``exotic'' Galilei group,
implemented as in (\ref{exoticrep}).
Then Noether's theorem allows us to derive the associated
conserved quantities~: if $L$ changes as
$\delta L=\p_{\alpha}K^{\alpha}$ under an
infinitesimal coordinate change $\delta \vx$, then
$
   \int\left(\frac{\delta L}{\delta(\p_{t}\phi)}\delta\phi
  +\delta\bar{\phi}\frac{\delta L}{\delta(\p_{t}\bar{\phi})}
     -K^{t}\right)d^2\vx
$
  is a constant of the motion.
  Using (\ref{exoticrep}) we get
\begin{equation}
     \begin{array}{ccc}
	{\cal H}&=\displaystyle\int\displaystyle\frac{\vp^2}{2m}
	\vert\phi\vert^2 d^{2}\vp\hfill
	&\hbox{energy}\hfill
	\\[5mm]
	{\cal P}_{j}&=\displaystyle\int p_{j}\vert\phi\vert^2 d^{2}\vp\hfill
	&\hbox{momenta}\hfill
	\\[5mm]
	{\cal J}&=\displaystyle\int\left(\displaystyle\frac{1}{2i}
	\epsilon_{jk}\big(\phi\displaystyle\frac{\p\bar{\phi}}{\p p_{j}}
	-
	\bar{\phi}\displaystyle\frac{\p \phi}{\p p_{j}}\big)p_{k}
	+s\vert\phi\vert^2\right)
	 d^{2}\vp\qquad\hfill
	&\hbox{angular momentum}
   	\hfill
	\\[5mm]
	{\cal G}_{j}&=m\displaystyle\int\left(\displaystyle\frac{1}{2i}
	\big(\phi\displaystyle\frac{\p \bar{\phi}}{\p p_{j}}
	-
	\bar{\phi}\displaystyle\frac{\p \phi}{\p p_{j}}\big)
	+
	\displaystyle\frac{\theta}{2}\epsilon_{jk}p_{k}\vert\phi\vert^2\right)
	d^{2}\vp\qquad\hfill
	&\hbox{boost}\hfill
	\\[5mm]
	{\cal M}&=m\displaystyle\int\displaystyle
	\vert\phi\vert^2 d^{2}\vp\hfill
	&\hbox{mass}\hfill
	\\[5mm]
	{\cal K}&=-m^2\theta\displaystyle\int\displaystyle
	\vert\phi\vert^2 d^{2}\vp\hfill
	&\hbox{exotic charge}\hfill\\
	\end{array}
\label{pconserved}
\end{equation}
These quantities are the expectation values of the
operators listed in (\ref{exGalop}) when the wave function is
normalized to $1$.
Consistently with our previous results, these conserved quantities,
with the exception of
the boosts, are standard.

The free Schr\"odinger equation (\ref{freeschreq}) is of the Hamiltonian form,
$
\p_{t}\phi=\Big\{\phi,{\cal H}\Big\}
$
  with the standard  Poisson bracket
\begin{equation}
     \big\{{\cal F}, {\cal G}\big\}=
     \frac{1}{i}\int\left(\frac
     {\delta{\cal F}}{\delta\phi}\frac{\delta{\cal G}}{\delta\bar{\phi}}
     -
     \frac{\delta{\cal G}}{\delta\phi}\frac{\delta{\cal F}}{\delta\bar{\phi}}
     \right)d^{2}\vp.
\label{pfieldPB}
\end{equation}
Then the quantities (\ref{pconserved}) are readily seen to close under
(\ref{pfieldPB})
into the exotic Galilean relations. In particular,
\begin{equation}
\big\{{\cal G}_{1},{\cal G}_{2}\big\}=-{\cal K},
\label{CCboostcr}
\end{equation}
cf. (\ref{boostcr}).

\section{Exotic symmetry of Moyal field theory}\label{MoyalFT}

Much recent work has been dedicated to
non-commutative field theory \cite{NCfield}, where the ordinary product
is replaced by the Moyal ``star'' product associated with
the non-commutative parameter $\theta$ \cite{Moyal},
\begin{equation}
\big(f\star g\big)(x_1, x_2)=\exp\left(i\frac{\theta}{2}\big(
\p_{x_1}\p_{y_2}-\p_{x_2}\p_{y_1}\big)\right)
f(x_1, x_2)g(y_1, y_2)\Big|_{\vx=\vec{y}}
\label{thetaMoyal}
\end{equation}
  associated with the  non-commutative parameter $\theta$.
Lozano et al. \cite{LMS}, consider, in particular,
a field theory inspired by ordinary, non-relativistic quantum mechanics.
We show now that such a field theory admits the
``exotic'' Galilean symmetry studied above.
  Let us indeed consider the free non-commutative field theory
  with the non-local Lagrangian
\begin{equation}
L_{NC}=
\frac{i}{2}\left(\bar{\psi}\star\p_t\psi-
\p_t\bar{\psi}\star\psi\right)
-
\frac{1}{2m}\vnabla\bar{\psi}\star\vnabla\psi.
\label{freeNClag}
\end{equation}

The Galilean invariance of this theory is obvious from the outset
since the Moyal product
can be ignored under integration, $\int f\star g\,d^2\vx=\int fg\,d^2\vx$.
(Alternatively, let us observe that the equation of motion
associated with (\ref{freeNClag})
is, despite the presence of the star product
in (\ref{freeNClag}), simply the free Schr\"odinger equation).
It is nonetheless useful to check the statement explicitly,
because the difference with the ordinary case
changes to the associated conserved
quantities. In fact, implementing a boost
  in the standard way, namely as
$\psi\to U_{\vb}\psi$,
\begin{eqnarray}
U_{\vb}\psi(x,t)=
e^{im(\vx\cdot\vb-\half b^2t)}\psi(\vx-\vb t,t),
\label{ordinaryboost}
\end{eqnarray}
changes $L$ into
\begin{eqnarray*}
\frac{i}{2}\left((e^{(-)}\bar{\psi})\star(e^{(+)}\p_t\psi)
-
(e^{(-)}\p_t\bar{\psi})\star(e^{(+)}\psi)
\right)
-
\frac{1}{2m}(e^{(-)}\vnabla\bar{\psi})\star
(e^{(+)}\vnabla\psi)
\end{eqnarray*}
taken at $(\vx'\!=\!\vx-\vb t, t)$,
where we used the shorthand
  $e^{(\pm)}=\exp[\pm im(\vx\cdot\vb+\half b^2t)]$.
In the commutative theory, this would be simply
$
\big[
\frac{i}{2}\big(\bar{\psi}\p_t\psi-\p_t\bar{\psi}\psi\big)
-\frac{1}{2m}\vert\vnabla\psi\vert^2\big]
= (\vx-\vb t,t)=L(\vx-\vb t,t).
$
In the noncommutative case, however,
  the Moyal product with the exponential factors results
in an additional shift of the argument,
\begin{equation}
(e^{(-)}f)\star(e^{(+)}g)(\vx)=
(f\star g)(\vx-\half m\theta\epsilon\vb)
\label{Lemma1}
\end{equation}
where $\epsilon=(\epsilon_{ij})$ is the totally antisymmetric matrix.
The NC Lagrangian
(\ref{freeNClag}) changes according
\begin{eqnarray}
L_{NC}(x,t)\to
L_{NC} (\vx-\vb t-\half m\theta\epsilon\vb,t).
\label{LNCchange}
\end{eqnarray}
Boosting is hence equivalent to a ``twis\-ted shift'',
so that the action $\displaystyle\int L_{NC}d^2\vx dt$ is invariant,
as expected.
  The  shift  yields, however, an additional term in the associated
conserved quantity. The definition
  (\ref{thetaMoyal}), together with Baker' formula \cite{Moyal},
\begin{equation}
     \big(f\star g\big)(x)=\frac{1}{(\pi\theta)^2}\int
     f(\vx')g(\vx'')e^{(2i/\theta)\Delta} d\vx'd\vx'',
     \qquad
     \Delta=(\vx'-\vx)\times(\vx''-\vx),
  \label{bakerprod}
\end{equation}
allow us to establish the relation
\begin{equation}
     \half\big[\bar{\psi}\star(\vx\psi)+
     (\vx\bar{\psi})\star\psi\big]=
     \vx\bar{\psi}\star\psi
     -\frac{\theta}{2}\epsilon\vj,
     \qquad
     \vj=\frac{1}{2i}
     \big(\bar{\psi}\star\vnabla\psi-\vnabla\bar{\psi}\star\psi\big).
     \label{Lemma2}
\end{equation}
Then for the conserved quantity associated with the boost we find
\begin{equation}
     {\cal G}_{i}=\int mx_{i}(\bar{\psi}\star\psi)\, d^{2}\vx
     -t{\cal P}_{i}
     +\half m\theta\epsilon_{ij}{\cal P}_{i},
     \label{xboost}
\end{equation}
  where
$
\vec{\displaystyle{\cal P}}=\displaystyle{\int}\vj\, d^{2}\vx=
\displaystyle{\int}(1/2i)
     \big(\bar{\psi}(\vnabla\psi)-(\vnabla\bar{\psi})\psi\big)\, d^{2}\vx
$
  is the conserved momentum, associated with the translational symmetry.
Note that  $\int x_{i}(\bar{\psi}\star\psi)\, d^{2}\vx\neq
\int x_{i}\vert\psi\vert^2)\, d^{2}\vx$, owing to the presence of
the coordinate $x_{i}$.

Similarly,  the energy, associated to the time translation, is
$
{\cal H}=\displaystyle
\int\frac{1}{2m}\vnabla\bar{\psi}\star\vnabla\psi d^{2}\vx=
\displaystyle\int({1}/{2m})\big\vert\vnabla\bar{\psi}\big\vert^2 d^{2}\vx.
$
  A rotation by $\varphi$ in the plane is implemented on the field according to
$
U_{\varphi}\psi(x)=e^{is\varphi}\psi(R\vx)
$
where $R=R_{\varphi}=e^{i\epsilon\varphi}$.
This leaves the free Lagrangian invariant and, using
\begin{equation}
      \half\epsilon_{ij}\big[\bar{\psi}\star(\p_{i}\psi x_{j})-
     (\p_{i}\bar{\psi}x_{j})\star\psi\big]
     =
     \vx\times\vj
     -\frac{\theta}{2}\vnabla\bar{\psi}\star
     \vnabla\psi,
     \label{Lemma3}
\end{equation}
we get the angular momentum involving both the exotic and the
spin terms,
\begin{equation}
     {\cal J}=
     \int\left(\vx\times\vec{\jmath}
     -\frac{\theta}{2}\vert\vnabla{\psi}\vert^2
     +s\ \vert\psi\vert^2
     \right)d^{2}\vx.
     \label{xangmom}
\end{equation}
Note that the new term due to the non-commutativity here
  is separately conserved, since it is proportional to the energy.

  Note that while ${\cal H}$, ${\cal P}$,
  ${\cal M}=m\!\displaystyle\int\!\vert\psi\vert^2\, d^2\vx$
  have the standard form,
the boost, ${\cal G}$, the angular momentum, ${\cal J}$, and
  the ``exotic'' central generator ${\cal K}=
  -m^2\theta\displaystyle\int\vert\psi\vert^2d^2\vx$
involve the non-commutative paremeter $\theta$.
These quantities, analogous to those found for a
classical particle in the non-commutative plane discussed in
\ref{NCQM}, span under the Poisson bracket,
(\ref{pfieldPB}) [with the integration variable $\vp$ replaced by $\vx$]
the ``exotic''  Galilei group.
Bracketing the boosts yields in particular once again
the ``exotic'' relation (\ref{CCboostcr}).

\section{Discussion}

For NCQM in the momentum space,
Section \ref{NCQM},
both the field-theoretical action (\ref{freepLag}) and the Poisson bracket,
(\ref{pfieldPB}) are conventional, and the only difference
with an ordinary (``non-exotic'') particle is the way boost acts on
the wave function, represented by the exponential factor
$\exp[im(\theta/2)\vb\times\vp]$ in Eq. (\ref{exoticrep}).
This is unlike as for a classical particle, where the
exotic structure appears in the Poisson bracket, (\ref{freeNCcomrel}),
while the Galilei group acts in the usual way.
  Remember, however, that the exotic structure could be made disappear
by redefining the position, namely introducing the ``Darboux''
variables $q_{i}=x_{i}+\frac{1}{2}\theta\varepsilon_{ij}\,p_j$
canonically
conjugate to the $p_{i}$. In terms of the $q_{i}$ and the $p_{i}$
we would recover the standard structure of
an ordinary particle -- but one upon which the Galilei group acts
in a non-standard way.

In NCFT considered in Section \ref{MoyalFT} instead, the action
(\ref{freeNClag})
involves the non-commuta\-tive parameter $\theta$ through the
Moyal star product, while the boosts act conventionally, cf.
(\ref{ordinaryboost}).
The action uses hence a non-local ``alternative Lagrangian''
for the free Schr\"odinger equation. It is precisely this modification
that opens the way for the ``exotic'' Galilean symmetry.

\goodbreak
\kikezd{Acknowledgement}
We are indebted to C. Duval and F. Schaposnik
for correspondence and enlightening discussions.

\section{Note added}

In calculating the commutator of the boosts above, an error was
comitted. Straightforward calculation shows indeed that
the components of the  ${\cal G}_{i}$ in (\ref{xboost})
in fact commute. This can also be understood by observing that
the  properties of the Moyal product allows us to
absorb the $\theta$-dependent term into the first integral so that
(\ref{xboost}) takes  the traditional form
\begin{equation}
     {\cal G}_{i}^0=\int mx_{i}\bar{\psi}\psi\, d^{2}\vx
     -t{\cal P}_{i}.
     \label{oldboost}
\end{equation}

Yet another explanation is obtained
by noting that the ``Moyal stars'' in (\ref{freeNClag})
can  be dropped. Owing to the ``integral property"
$\int f\star g=\int fg$  the Lagrangian (\ref{freeNClag}) is in fact
equivalent to the standard free expression whose associated conserved boost
is (\ref{oldboost}).

Our theorem is, nevertheless, correct~: it is in fact enough to
implement the boosts by inserting a Moyal star into (\ref{ordinaryboost})
i. e. to consider rather
\begin{eqnarray}
U^\star\psi(x,t)=
e^{im(\vx\cdot\vb-\half b^2t)}\star\psi(\vx-\vb t,t).
\label{starboost}
\end{eqnarray}
This novel type of action is still a symmetry and yields, instead of
(\ref{oldboost}),
\begin{equation}
     {\cal G}_{i}^*=m\int\!d^{2}\vx x_{i}\bar{\psi}\psi
     -t{\cal P}_{i}-\frac{\theta}{2}\epsilon_{ij}{\cal P}_{j}
     \label{newboost}
\end{equation}
which does indeed satisfy the exotic commutation relations
\begin{equation}
     \big\{{\cal G}_{1}, {\cal G}_{2}\big\}=\theta\int\!d^2\vx\bar{\psi}\psi.
\end{equation}
This corrects an error in \cite{DHspin}.
See \cite{HMS} for further details.

     We are indebted to Professor P. Stichel for calling our attention to
this point.
\goodbreak


Similarly, the 2-dimensional version of L\'evy-Leblond's
``non-relativistic Dirac equation'' \cite{LLCMP} equation can be
considered \cite{Hagenexotic, DHspin}.
Let $\Psi$ denote indeed a two-component spinor, and
consider the Lagrange density
\begin{equation}
     L=\Im\left\{\Psi^\dagger\big(\Sigma_{t}\p_{t}
     -\vec{\Sigma}\cdot\vnabla+i\Sigma_{s}\big)\Psi\right\}
     \label{2LLlag}
\end{equation}
where
\begin{equation}
     \Sigma_{t}=\half(1+\sigma_{3}),
     \qquad
     \Sigma_{i}=\sigma_{i}
     \quad(i=1,2),\quad
     \Sigma_{s}=(1-\sigma_{3}).
     \label{Sigma}
\end{equation}
Setting $\Psi=\left(\begin{array}{ll}\Phi\\
\chi\end{array}\right)$ yields the planar version of the
L\'evy-Leblond  equation \cite{LLCMP},
\begin{equation}
     \begin{array}{ccc}
	(\p_{1}+i\p_{2})\Phi+2i\chi\hfill&=&0
	\\[6pt]
	\p_{t}-(\p_{1}-i\p_{2})\chi\hfill&=&0
	\label{2DLLeq}
     \end{array}
\end{equation}

Implementing a boost conventionally as \cite{LLCMP, DHP}
\begin{equation}
     U\Psi(\vx,t)=\left(\begin{array}{cc}
     1&0\\-\half(b_{1}+ib_{2})&1
     \end{array}\right)
     e^{i(\vb\cdot\vx-t\vb^2/2)}\Psi(\vx-t\vb,t)
     \label{spinCboost}
\end{equation}
or infinitesimally, as
\begin{equation}
     \begin{array}{ll}
     \delta\Phi=i\vb\cdot\vx\Phi-t\vb\cdot\vnabla\Phi
     \\[6pt]
     \delta\chi=\half(b_{1}+ib_{2})\Phi
     +i\vb\cdot\vx\chi-t\vb\cdot\vnabla\chi
     \end{array}
     \label{spininfCboost}
\end{equation}
we find that the LL equations (\ref{2DLLeq}) remain satisfied,
establishing the Galilean symmetry.

The associated conserved boost components, again (\ref{oldboost})
[with the upper component, $\Phi$, replacing the scalar field,
$\psi$], commute, as observed by L\'evy-Leblond 35 years
ago and stressed recently by Hagen \cite{Hagenexotic}.

Inserting a Moyal ``star'' into the free Lagrangian (\ref{2LLlag})
would yield an equivalent (non-local) Lagrangian for which the conventional
implentation is still a symmetry; the yields the {\it same} boost
generators with commuting components. (This corrects another false statement
made in \cite{DHspin}).
This is also seen by observing, as above, that formula (2.13) in that paper,
although correct, can again be transformed into the conventional form
using the above-mentioned properties of the Moyal product.

The statement made in \cite{ DHspin} is still correct:
it is enough to change the conventional implementation
once again by inserting the
Moyal product i. e. to consider rather
\begin{equation}
     U^*\Psi(\vx,t)=\left(\begin{array}{cc}
     1&0\\-\half(b_{1}+ib_{2})&1
     \end{array}\right)
     e^{i(\vb\cdot\vx-t\vb^2/2)}\star\Psi(\vx-t\vb,t)
     \label{spinNCboost}
\end{equation}
or infinitesimally
\begin{equation}
     \begin{array}{ll}
     \delta^*\Phi=(i\vb\cdot\vx)\star\Phi-t\vb\cdot\vnabla\Phi
     \\[6pt]
     \delta^*\chi=\half(b_{1}+ib_{2})\Phi
     +(i\vb\cdot\vx)\star\chi-t\vb\cdot\vnabla\chi,
     \end{array}
     \label{spininfNCboost}
\end{equation}
to get the  ``Moyal-boost'' (\ref{newboost}) [with $\psi\to\Phi$]
which has noncommuting components.

Further generalization will be found elsewhere \cite{HMS2}.


\end{document}